\documentclass[referee]{aa} % for a referee version
\newcommand{\ltapprox}{\raisebox{-0.5ex}{$\,\stackrel{<}{\scriptstyle\sim}\,$}}

\begin{document}
\input psfig.tex
 
\title{Homogeneity of early--type galaxies across clusters}

\author{S. Andreon}
%\inst{1}
\institute{INAF--Osservatorio Astronomico di Brera, Milano, Italy}

% \and Department of Astronomy, Caltech, USA}

%\and R. Ellis\inst{2}

\titlerunning{}

\date{running version 1.1}
 
\abstract{We studied the homogeneity, across clusters, of the color of the red
sequence (the intercept of the color--magnitude relation) of 158 clusters and groups
detected in the Early Data Release of the Sloan Digital Sky Survey (EDR--SDSS) in the
redshift range $0.06<z<0.34$. We found a high degree of homogeneity: the color of the
red sequence shows an intrinsic scatter of 0.02 mag across clusters, suggesting that
either galaxies on the red sequence formed a long time ago ($z>2$) or else their
star formation is universally delayed with preservation of a small spread in age
formation. The latter possibility is ruled out by the mere existence of galaxies at
high redshift. While the old age of ellipticals was already been claimed for a small
heterogeneous collection of clusters, most of which are rich ones, we found that it
holds for ten to one hundred large sample, representative of all clusters and groups
detected on the EDR--SDSS. Hence we suggest the possible universality of the color of
the galaxies on the red sequence. Furthermore, the sample includes a large number of
very poor clusters (also called groups), not studied in previous works, for which the
hierarchical and monolithic scenarios of elliptical formation predict different colors
for the brightest ellipticals. The observed red sequence color does not depend on
cluster/group richness at a level of 0.02 mag, while a $\sim 0.23$ mag effect is expected
according to the hierarchical prediction. Therefore, the stellar population of red
sequence galaxies is similar in clusters and groups, in spite of different halo
histories. Finally, since the observed rest--frame color of the red sequence does not
depend on environment and redshift, it can be used as a distance indicator, with an
error $\sigma_z=0.018$, a few time better than the precision achieved by other
photometric redshift estimates and twice better
than the precision of the Fundamental
Plane for a single galaxy at the median redshift of the EDR--SDSS.
 }
 \maketitle %\newpage

\section{Introduction}

The existence of a relation between color and magnitude for early type galaxies is
known since a long time (e.g. Sandage \& Visvanathan 1978). This relation, known as
color--magnitude relation, implies a link between the mass of the stellar population
and its age or metallicity, with the latter being the presently favored explication,
because high redshift clusters have bluer (rest--frame) color--magnitude relations with a small
scatter (Ellis et al. 1997, Kodama \& Arimoto 1997, Stanford, Eisenhardt \& Dickinson
1998; Kodama 1998). The color--magnitude relation, often called red sequence, of
different clusters shows an homogeneity across clusters, in the sense that the
(rest--frame) color of the red sequence is similar for the presently studied clusters
(e.g. Garilli et al. 1996), although the sample studied thus far presents several
limitations described in the next paragraph. A color uniformity, inside each cluster
or across different clusters, implies that  stellar populations are similar, or old
enough that differences  induced by a spread in formation ages are dumped out. 

At first sight, the early formation of the bulk of stars in cluster galaxies  
seems a problem for hierarchical scenarios of galaxy formation. However,
while star formation occurs late in such a models, it is pushed back to early times in
clusters (Kauffmann 1996). Hence the conclusion that  in clusters the bulk of stars in
galaxies is homogeneous and red  is not in contradiction with the hierarchical
scenario. A natural consequence of hierarchical models would be that the evolution of
early--type galaxies should depend on their environment: merging history are likely to
have varied from cluster to groups, and also across groups because they are formed by
the coalescence of a tiny number of halos.

From the observational side, Bower, Lucey \& Ellis (1992), Aragon-Salamanca et al.
(1993), Ellis et al. (1997), Stanford, Eisenhardt \& Dickinson (1998), Kodama et al.
(1998), all studied the homogeneity of the color of the red sequence in
2, 10, 3, 19, 17 clusters, respectively, with the latter paper drawing
data from previous papers and performing a comprehensive analysis of the same
observational data. All but two studied clusters are rich or very rich. Rich clusters
are the ones for which a hierarchical scenario and the usual monolithic scenario for
elliptical formation (Eggen, Lynden-Bell \& Sandage 1962) give the same prediction
about early--type color and dispersion around the color--magnitude relation (Kauffmann
\& Charlot 1998) and hence rich clusters cannot be used to discriminate between the
two scenario. Finally, the  clusters studied thus far are mostly at intermediate
redshift ($z>0.3$), and therefore our local universe is mostly unexplored in this
sense.

To summarize, the sample studied thus far is an heterogeneous collection of
clusters, and not a representative sample of clusters in the Universe.
Hence, what we know on the homogeneity of ellipticals is mainly based
on a restricted number of rich clusters, mostly at high redshift,
selected in an uncontrolled way.

In this paper we present mainly an observational effort to 
study a sample:

-that is larger (by a factor 10 to 100) than previous works;

-that is a representative sample of local clusters and groups  detected
in the Sloan Digital Sky Survey;

-that spans a broad range in richness, including groups, hence allowing us
to discriminate between monolithic and hierarchical scenarios of elliptical
formation.

\medskip
Section 2 presents the cluster sample and the data used. Observational
results are presented in section 3 and are summarized and discussed
in section 4. In section 5 we model the star formation history
of the clusters and compare it with the observations. Finally, section 6
summarizes the conclusions.

All along the paper we use robust statistics, even if we refrain ourselves to
remind it. In place of the mean we use the median, and in place of the dispersion
we use 1.47 times the interquartile range (the coefficient is appropriate for a
Gaussian distribution), but we still call them ``mean" and ``dispersion",
respectively.

No cosmological model is needed in the analysis of Section 2 and 3, while a
``concordance" model ($H_0=70$ km s$^{-1}$ Mpc$^{-1}$, $\Omega_m=0.3$ and 
$\Omega_{\Lambda}=0.7$) is used for conversion from times to redshift in
Section 5. In the course of the Section 4 absolute quantities (such as absolute
magnitudes or metric radii) are mentioned using $H_0=50$ km s$^{-1}$ Mpc$^{-1}$
for allowing an immediate comparison with published results derived in 
the ``standard" model, but the important quantities
are the apparent ones, that are independent on the cosmological parameters.

\section{Data \& Method}

The aim of this section is to measure the color of the red sequence in
a large sample of groups and clusters. To this aim, we present the
galaxy catalog, the detection algorithm used to detect clusters and
groups, the way we measure redshifts of our clusters and groups, and
the biases induced by the detection algorithm and by the redshift determination,
with particular emphasis on those potentially discarding  
clusters having a red sequence with color different from the average.

\subsection{The galaxy catalog \& the cluster detection algorithm}

We used the Early Data Release of the Sloan Digital Sky Survey (EDR--SDSS) catalog
(Stoughton et al. 2002), of which we considered a sub--area of 294.6 deg$^2$ having a 
rectangular shape. The galaxy catalog have been filtered out by residual stellar contamination and
by saturated objects, as described in Andreon (2003). $g'-r'$ color has been computed
using fiber (3 arcsec aperture) magnitudes. All magnitudes and colors have been
corrected for Galactic absorption, using the extinction values listed in the EDR--SDSS
database. Only galaxies brighter than $r_{Petro}=21.5$ mag are considered ($M^*+2$
at $z=0.3$ for unevolving early--type galaxies, assuming the $M^*$ value quoted in
Blanton et al. 2001).

%We note here that 
%colors computed through aperture magnitudes are robust, because computed on the 
%same aperture, and because $g'$ and $r'$ observations are almost simultaneous, thus sharing 
%very similar seeing and atmosphere conditions.

Clusters of galaxies have been detected by a method inspired to, but fundamentally
different from, Gladders \& Yee
(2000) and fully described in Andreon (2003). Shortly, the method takes
advantage from the observed fact that most of galaxies in clusters shares
similar colors, while background galaxies have a variety of colors, both
because they are spread over a larger redshift range and because the field
population is more variegated in color than the cluster one, even at a fixed
redshift. In practice, the method looks at spatially localized galaxy
overdensities of similar color. It works because the filtering in color reduces
the background galaxies (which have a variety of colors) without completing
discarding the cluster contribution. Being the cluster search almost in three
dimensions (color plus two spatial directions), projection effects that plagued
previous two dimensional searches are strongly reduced.

For each color slice, whose central color is made to vary in order to
detect galaxy overdensities at different colors, a multiscale detection
algorithm is applied to the map of the numerical density of galaxies. By a
friend--of--friend algorithm, detections at different scales and colors are
then grouped together when they overlap in sky and in color. The ``winner"
detection is the one that maximizes the numerical density of galaxies.

We applied the cluster detection algorithm, using a 0.1 mag wide strip in color,
stepped by half its width, and 2.5, 5 and 10 arcmin boxes, stepped by half their width.
The catalog used in this paper is drawn from the one formed
by the 643 statistically most significant detections, defined here as those having a
probability smaller than $10^{-6}$ to be statistical fluctuations.  The size
of the ``winner" detection has the mode at $\sim 500$ kpc, 
the minimum at $\sim200$ kpc and a tail up to 4 Mpc.

%larger than $0.99999$  to be real clusters or groups.

Our detection algorithm does not bias the catalog against clusters  having
galaxies with abnormal color (e.g. with a color--magnitude relation having an
usual red or blue color),  simply because the detection is performed at all
colors without any constrain on the expected galaxy color at a given redshift. If
clusters having their color--magnitude relation bluer or redder than the average
exist, they have the same chance (or very similar ones) to be detected as normal
clusters. This point is illustrated in Fig 1. Fig. 1 shows, for 16 randomly
selected clusters (including some clusters outside the redshift window considered
in the present work) the color distribution of the galaxies in the cluster line
of sight (solid histogram) and in the control field (dashed histogram), taken to
be a control area more than 100 deg$^2$  wide.  A clear spike (the cluster
contribution) is always present at the cluster location. Since the background
color distribution is quite flat, moving the cluster contribution by, say, 0.2
mag toward the red or the blue, does not change the cluster detectability,
because the contrast between cluster and background stays approximately constant.
Quantitatively, simulations shows that 93 \% of the actually detected
clusters are detected at $10^{-6}$ probability threshold even if the background
distribution is made bluer or redder by 0.4 mag (that is mathematically equivalent
to move cluster galaxies by $\pm0.4$ mag in color).  In other terms, clusters
with galaxies too blue or too red than average (by a reasonable quantity, say
0.2--0.4 mag), are not lost because the probability of their detection is
independent on color for small (0.2 to 0.4 mag) color variations. I.e. clusters
with red sequences of average color, bluer or redded than average, all have very
similar probability to be detected. Therefore, the detected clusters are not
biased against or contrary clusters having the ``right" colors. 

{We have therefore shown that the detection probability is approximately constant
if the cluster color distribution is shifted toward the blue or the red by a reasonable
quantity. Of course, the detection probability depends on the shape and width of the
color distribution of cluster members: the wider and smoother the distribution is, the
lower is the probability that the cluster is detected.  Qualitatively, the
detectability of a cluster (or group) whose red sequence galaxies are significantly
bluer than the usual color of early--type galaxies (the prediction of the hierarchical
scenario) is a delicate balance of several factors, such as the sharpness and width of
the color distribution, Poissonian fluctuations, the possibility of an age--metallicity
degeneracy.  The precise calculation of this balance of outside the aims of this work, 
because this paper only consider color shifts (red sequences bluer or redder than
average). However, let us consider just two cases. According to Frei \& Gunn
(1995), the available color range (for galaxies of all spectrophotometric types later
than Irr) is only 0.2 mag at $z=0.2$ ($z=0$) if the red sequence is bluer than average
by 0.4 (0.2) mag.  In these cases,
the maximal color range is only twice larger than the color width used
in the detection. Therefore, groups at these redshifts twice richer than
the poorest detected are detected, whatever is the shape of their color distribution. 
In realistic cases, the factor two is an overestimate.

Although is not completely relevant here, it should emphasized that many of the
detected clusters would be detected even without their red sequence galaxies,
hence by surreptitiously reducing the group richness. For
example, the cluster cl27 in Figure 1 is detected not only at $g'-r'=1.15$ mag
(the color of the red sequence), but also at colors as blue as  $g'-r'=1.00$
(i.e. $0.95<g'-r'<1.05$ mag), where the contribution of red sequence galaxies is
deemed to be  negligible. Cl196, also shown in Fig 1, is another example, at
larger redshift. Just a few galaxies with similar colors (not necessarily red
sequence galaxies) are enough to trigger the cluster detection. For example 6 galaxies
of color $0.95<g'-r'<1.05$ mag and in a area of 6.25 arcmin$^2$ are enough to
trigger a cluster detection, no matter whether these galaxies are early--type
galaxies on the red sequence at $z\sim 0.13$ or intrinsically bluer galaxies at
higher redshift. Therefore, clusters detected by their blue component have a red
sequence with a color unbiased for one more reason: these clusters would be
detected independently on the existence (and therefore the color) of the red
sequence. 

Finally, the
simple existence in our catalog of very low richness clusters, that are known 
to be dominated by late--type galaxies and hence to contain just a few red
galaxies, suggests that the cluster detection algorithm is not strongly biased
against clusters dominated by late--type galaxies.

Gladders \& Yee (2000) follow a different path, apparently similar
but in fact fundamentally different. They
search for clusters, at a given redshift, with the ``right" angular size,
with a red sequence of the ``right" color
and with a luminosity function of the ``right" shape and characteristic
magnitude. In their search, the relation between redshift, luminosity
early--type galaxy color and cluster angular size is imposed in the detection
algorithm, hence enhancing the detection of objects that look like
the expectations.
However, the detection probability for clusters that differ from
our expectations (in the
color of the red sequence, for example) is reduced. 

In order not to bias our sample against ``unusual" objects we perform:

1) a multi--scale detection (with scale unrelated to distance),
without a weight that maximize the detection for the clusters
having the ``right" radial profile;

2) we don't use the information provided by the magnitude, but we
use the colors only;

3) we don't relate distance and color via a model for the spectrum
of the detected galaxy overdensity: overdensities of a given
color can be at whatever redshift.

Hence, our search is sensitive to a broader range of type of clusters, not
discarding {\it ab initio} those that do not satisfy our priors, at the
price, of course, of a reduced detection efficiency on clusters having 
common properties. In particular, clusters detected by our method can have
red sequences of whatever color, without biases induced by selection effects
imposed by hand in the software detection algorithm.

We re--emphasize here that the optical selection is not, in general, the best
way to select the objects of which we want to study the evolution (Andreon \&
Ettori 1999). However, in this specific case, the concern does not
hold because we are not artificially imposing a selection effect that
mimics the observed behavior: we are not discarding ab initio objects
that do not satisfy the possible
universality of the color of the red sequence.

In the cluster detection phase we neglect the slope of the color--magnitude
relation in order not to bias our cluster sample toward clusters having the
``right" color--magnitude slope.

What type of cluster are we sampling? By looking at the Abell (1958) catalog in
the EDR--SDSS area, we found that there are Abell (1958) clusters of richness $R$
from 0 to 3, plus numerous clusters that do not have
counterpart in the Abell catalog and have $z<0.1$. Since the Abell catalog is
claimed to be statistically complete up to $z\sim0.1$ for $R \geq 1$ clusters (Abell
1958), our clusters with $z<0.1$ and without Abell counterparts are missing in
the Abell catalog (mainly) because they are
too poor. Therefore, our sample includes
several examples of $R \leq 0$ clusters, i.e. groups of less than 50 members ($R=0$) or
30 members ($R<0$), according to the Abell definition of richness. This type of
groups are those specifically considered in Kauffmann (1996) in their
estimate of the color of the red sequence in a hierarchical scenario. 

 In the studied area and
in the considered redshift range ($0.06<z<0.34$), none of the Abell clusters with at least 3
concordant redshifts, or with some measured X-ray emission is missed by our algorithm.
However,
the comparison of our cluster catalog with the Abell (1958) one does not provide significant
informations about our cluster detection efficiency, given scarce available data (redshift,
assessment of the reality of the Abell detection) and the small size of starting sample (49
clusters in our area, including Abell clusters without known redshift).

\subsection{Color, Redshift and color biases}

Almost all detected clusters do not have a known redshift. The EDR--SDSS includes
redshift for galaxies, but not for structures (groups and clusters). The wealth of
information provided by the EDS-SDSS allows different paths for deriving the
cluster redshift. Our redshift determination need to be independent on galaxy colors (in
order not to introduce an artificial bias/link between redshift and color, that is the
argument under study) and its precision should be similar for both very rich and very
poor clusters (in order to make simple the statistical analysis). These requirements
discard the possibility to use photometric redshifts and force us to use the
spectroscopic catalog included in the EDR--SDSS. The obvious path for the redshift, 
i.e. looking for a peak in the redshift distribution is likely unsuccessful,
especially for poor clusters or very distant ones (the largest peak would be
the median redshift of the survey, not the cluster redshift). Therefore,
we took a much simpler approach: the cluster redshift is the redshift of one
galaxy, as specified below.

We make two approximations, that will be checked in a while:

1) the color of the red sequence is the {\it color} of any galaxy with measured
redshift within the color strip (0.1 mag wide) and area of cluster 
best detection.

2) the cluster redshift is the {\it redshift} of the galaxy defining
the color of the red sequence. 

Approximation n. 1 introduces a very small error: one single galaxy locates the
red sequence within 0.038 mag at worst (i.e. the dispersion of an uniform
deviate between -0.05 and 0.05 mag). If the intrinsic scatter around
the color--magnitude relation is 0.035 mag, as observed for Coma and Virgo
clusters in $U-V$ (Bower, Lucey \& Ellis 1992), we are locating the red sequence
with an error of 0.038 instead of 0.035/$\sqrt{n}$, where $n$ is the number of
galaxies eventually used to locate the sequence. Since there are, very often,
only a few galaxies on the cluster center of our (mostly poor and very poor)
clusters, the use of a single galaxy to locate the sequence do not degrade our
results too much. This choice also simplifies the statistical analysis, because
all red sequences have similar errors on color, independently on the cluster
richness.

Approximation n. 2 has two drawbacks: first, the cluster redshift cannot be
measured with a better precision than the typical velocity dispersion
of galaxies in clusters, and,
second, some {\it assigned} cluster redshifts can be completely
wrong when the galaxy used to measure the redshift is
an interloper. Hence, approximation n. 2 inflates the
{\it measured} color dispersion of the red sequence and introduces
outliers in their color distribution.

Figure 2 shows the error introduced by measuring the cluster redshift using a
single galaxy, by histogramming the 95 values of the individual
$\Delta z$ that we computed
for clusters with two or more galaxies with known redshift. The distribution
peaks to $z=0.000$ (i.e. no systematics are there) and the width of the
distribution is 0.0033 in
$z$ (or about 1000 km s$^{-1}$). Given the observed relation between color
and redshift (Figure 3, discussed later) the cluster velocity dispersion
enlarges the dispersion in
the {\it rest--frame} color by $0.01/\sqrt{2}$ mag, a
negligible quantity. 

Most importantly, only 17 \% of the clusters are outliers, i.e. their 
$\Delta z$ is off, by more than 3 $\sigma$ (i.e. by about 0.01).

There is a second method to derive the percentage of wrong redshift assignations.
In the area of best detection, there are 2726 galaxies within the 0.1 mag strip, 
and, statistically, only 488 (i.e. $\sim 18$\%) should be
background galaxies\footnote{The number of background galaxies is given by the
observed cluster area times the observed average density of galaxies  
within the 0.1 mag strip
over the whole EDR--SDSS area.}. Galaxies with redshift in the
EDR--SDSS database are a subsample of the 2726 galaxies, skewed toward cluster
members for two reasons. First of all, the main EDR--SDSS survey targets only bright
($r' \ltapprox 17.77$ mag) galaxies, while the number computed thus far are for $r'<21.0$
mag. At bright magnitudes the cluster contrast with respect to the background is
higher, by a factor of several, than over the whole studied magnitude range.
Hence, the contamination is likely 5 to 10 \%, or less. Second, fainter
galaxies with measured redshift belong to the ``Luminous Red Galaxy" sample,
that targets the most luminous and red galaxies in the Universe, common in
clusters and rare in the field. This selection further reduces the expected
contamination. A posteriori, only few red sequences are much redder or bluer
than the average, implying that wrong cluster assignations are rare, as we
will show in section 3.1 by a third method. 

To conclude, one redshift
is enough to determine the cluster redshift if it comes from a red sequence
galaxy, 90 \% of times. Robust statistics, that we adopt, overcomes the effect
of a few mis--assigned redshifts. On the other end, we cannot firmly exclude
the existence of very rare clusters whose red sequence has a color very different
from the remain of the sample.

To summarize, the two drawbacks of assumption n. 2 have a negligible impact or
concern a minority of clusters, and the latter can be deal with by using
robust statistics.

Out of 634 clusters, 158 have a known redshift in the range $0.06<z<0.34$. 
In the photometric redshift range $0.06<z<0.30$, where
photometric redshifts are accurate (see Sect. 3.7), there are 370 clusters, of
which 127 have a known redshift. Therefore, the studied sample is
about one third of the clusters detected in the EDR--SDSS, in the same redshift range.

One more technical detail should be now mentioned: when a cluster has $n>1$ galaxies with
redshift, approximations 1) and 2) need one more specification: which one, among the galaxies
with redshift, is the galaxy that defines the color and redshift of the cluster. The analysis
presented in this paper is performed in two different cases: a) taking the galaxy nearest to the
cluster center or b) taking all the galaxies, i.e. duplicating the cluster $n$ times, once per
galaxy with $z$. Since the results are independent on such an assumption, we quote only results
for case b).

\subsection{Color biases induced by spectroscopy}

The EDR--SDSS catalog includes a galaxy spectroscopic sample composed by two
subsamples. A first one is an usual flux--limited sample ($r' \ltapprox 17.77$) mag (Strauss et
al. 2002). The second one, the ``Luminous Red Galaxy" (Eisenstein et al. 2001) sample 
is aimed to get redshifts for an absolute magnitude limited sample of intrinsically
red galaxies. Therefore, the sample of galaxies in the spectroscopic database has a
characteristic feature: at $z<0.2$ the galaxy luminosity
function is sampled with variable depth, from $M_r\sim-21$ mag to $M_r\sim-23$ mag. 
At $z>0.25$ the ``Luminous Red Galaxy" sample dominates, the limiting magnitude does
not longer increases, and only galaxies brighter than $M_r\sim -22.5 -23.0$ mag are
included. These features are shared by the subsample of the EDR--SDSS
spectroscopic database that we consider in the following.

At $z<0.25$ there is no relationship between the color of the galaxies on the red
sequence and their presence in the spectroscopic database, simply because the color of
the galaxies in the main spectroscopic sample  are not used for the spectroscopic
selection. Therefore, the subsample of clusters having redshift drawn from the main
spectroscopic sample is a random (and representative) subsample of the whole sample of
clusters listed in the original cluster catalog. Neither the average nor the scatter
of the color of red sequence are hence affected by selection effects induced by the
requirement that the cluster redshift should be known. The same should be verified to
hold too for the high redshift ($z>0.25$) subsample, which is, instead, color
selected, because the ``Luminous Red Galaxy" sample put fibers preferentially on red
galaxies, hence potentially excluding a priori clusters with a blue red sequence, if
they exist. The possible lack of clusters in one tail of the distribution (the blue
side)  could hence reduce the {\it observed} scatter and introduce a systematic bias
between the {\it observed} average colors and the true average color. However,
Eisenstein et al. (2001) show that the density of the Red Luminous Galaxies is
constant up to $z=0.35$. Their Figure 3 shows that the cuts used for the target
selection is about 0.23 mag away from the expected color of Es (please note that the
track shown in their Figure 3 is made bluer by 0.08 mag, as explained in their
Appendix 2), and hence the selection only affects galaxies too blue to be of interest
here. Therefore, at the largest studied $z$ too, the target selection is
color--independent for red sequence galaxies, and both the average color and the
scatter are safe.

%Since there is about one galaxy per cluster, on average, in our sample, the
%color dispersion around the red sequence in each individual cluster cannot be
%{\it directly} computed. However, the scatter of the color of the red sequence
%among cluster can be computed, because one single galaxy locate the
%red sequence within 0.038 mag at worst.

\section{Results}

\subsection{Main trend}

Figure 3 shows the $g'-r'$ color as a function of redshift, for $0.06<z<0.34$.
At each redshift a finite and small dispersion is found (dashed error bars).
Furthermore, there is a clear trend for redder colors at high redshift, as
expected, due to the differential k-corrections between the two filters.
Therefore, any measure of the color dispersion should be reduced to the
rest--frame.

Figure 3 shows also errors on the mean color at each redshift (small solid error
bars) and the expected track of an unevolved (spectro--photometric)
elliptical (continuous line). The
error on the mean (median) is assumed to be equal to the dispersion divided by
$\sqrt{N*0.8}$ in order to approximately take into account the presence of
outliers and the reduced efficiency of the median with respect to the mean. The
track has been computed by using the elliptical spectrum listed in Coleman, Wu
\& Weedman (1980), adopting the filter response of EDR--SDSS filters, convolved
with the CCD quantum efficiency and atmosphere transmission, as listed in the
EDR--SDSS web page\footnote{
http://archive.stsci.edu/sdss/documents/response.dat}. The rest--frame color of
ellipticals is taken to be $g'-r'=0.74$ mag, (Frei \& Gunn 1994). Our
k--corrections are almost identical to those listed in Frei \& Gunn (1994) and
Fukugita, Shimasaku, \& Ichikawa (1995), but we have a dense sampling in $z$,
absent in the cited papers. The continuous line is not a fit to the data, being
no free parameters. Qualitatively, the agreement is very good: the observed
color change is the one expected
assuming no evolution and an elliptical spectrum for red sequence galaxies. The
agreement is even surprising, considering the claim that there are residual
(admittedly small) errors in the precise photometric calibration of the
EDR--SDSS data (Stoughton et al. 2002). Given the observed agreement between
observed and expected colors, residual calibration errors should almost cancel
out in the $g'-r'$ aperture color.

The agreement between the observed color and the non--evolving E track is expected, because the
passive evolving track (computed by using Bruzual \& Charlot (1993) models), and the
not--evolving track are expected to be almost identical (passive evolution accounts for $\sim
0.02$ mag of blueing at $z=0.25$). 

%The flattening of the color track at $z\sim3$ happens because
%the decrease of the $g'$ flux and is almost balanced by the passage of the
%$r'$ filter through the CH G--band.

The dotted line shows an empirical interpolation, as provided by an artificial
neural network trained on the individual $(z,g'-r')$ pairs. The two curves are
identical, showing that in order to derive the scatter around the track it
makes no difference to assume a model for the redshift dependence of $g'-r'$
color or to empirically derive the latter directly from the data themselves.

%Second, by tracking the red sequence at different
%redshifts, on does not trace the continuous evolution of a single collection of
%galaxies, but instead selects only the oldest galaxies at any given redshift,
%which thus mimic a passive evolutionary behavior or even a not evolving sample 
%(Stanford, Eisenhardt \& Dickinson 1998; van Dokkum \& Franx 2001). Non so
%se quest'ultimo si applica a me.!!!!!!!!!!!!!!!!

The median absolute difference between the track and the points is 0.01 mag, and
therefore the track is an {\it accurate} description of the data. According the
$\chi^2$ test, the E track is an {\it acceptable} description for the blueing of
the red sequence, since the reduced $\chi^2$ is $\chi_{\nu}^2=1.68$ for 11 degree
of freedom, and the track can be rejected at the 90--95 \% confidence level only. 

There are small deviations from the track
and sophisticate statistical tests that use the individual 253 data points and
that look for correlated deviations from the track in localized redshift ranges
(focus your attention, for example, on the three points below the track at $z\sim
0.25$, each one being the median of about 20 individual data points) can reject
the track as acceptable description of the underlining individual data points at a
large confidence level. There are two reasons for expecting these (admittedly
small) deviations. First of all, the track gives the expected color of an elliptical
galaxy of a fixed {\it absolute} magnitude. In our sample, instead, the absolute
magnitude admits all values that allow the galaxies to be included in the
spectroscopic sample. The average absolute magnitude of
galaxies in the spectroscopic database depends on redshift, because the
spectroscopic sample is largely a flux limited sample (Section 2.3). Because of the
color--magnitude relation, at large redshifts the sample is dominated by brighter
and therefore redder than average early--type galaxies, while at low redshifts
the sample is dominated by fainter and therefore bluer than average early--type
galaxies. Second, colors are measured in a fixed aperture in the observer frame
(a 3 arcsec aperture), not in a fixed aperture in the galaxy frame as implicitly
assumed in the computation of the track. Because of existence of 
color gradients in early--type galaxies and the use of a fixed aperture in the
observer frame, distant galaxies appear bluer (because aperture is larger in their
rest--frame), than nearby ones, all the remaining kept fixed. 

Therefore, the color of the objects should be reduced to a fixed absolute
magnitude, by adopting a slope for the color--magnitude relation and a
correction for the color gradients inside the galaxies. Both these informations
(slope of the color--magnitude relation, amplitude of the color gradients, and
eventually their dependence on the absolute magnitude) are unknown at the
present time for the EDR--SDSS photometric system. We can, however, take an
empirical approach, by reducing the color of the galaxies to those of
$M_r=-22.5$ mag galaxies by adopting a correction 
that is similar to the color--magnitude relation: 

$(g'-r')_{corr}=(g'-r')+\alpha(M_r+22.5)$ 

where $\alpha$ is a parameter to be determined from the observations, and
$M_r$ is the absolute magnitude of the considered galaxy.
Adopting $\alpha=-0.02$, we are able to remove the dependence of color
residuals from redshift, dependence pointed out by correlated deviations
from the track in localized redshift ranges. This correction
is of the order of 0.01 mag. Figure 4 shows the
{\it corrected} $g'-r'$ color, as a function of redshift. 

Since the average color of the red sequence is well described by our
synthetic track, at the 0.01 mag (median absolute
deviation between the track and the data points), and residuals do not
longer correlate with redshift, we can remove the redshift dependence of
the red sequence color fairly accurately. Because of the very tiny
mismatch between the data points and the track, the dispersion 
(robust dispersion of the absolute deviations) in color of
the red sequences is inflated by 0.015 mag, but without dependence on $z$. 

\subsection{Scatter of red sequence}

Figure 5 shows color residuals from the E track. The (robust) dispersion,
i.e. the observed scatter in color of the red sequence across clusters, is
0.054 mag, as listed in Table 1. We remind that the galaxies entering in
this plot (and following ones) are only those having a spectroscopic
redshift (158 clusters sampled with 253 galaxies). Their absolute magnitude
distribution is described in section 2.3. 

We checked if part of this scatter is due to systematic effects,
such as color drift across the survey. We was unable to
find any trend between color residuals and observing run,
CCD id, right ascension or declination, hence supporting the
non--systematic nature of the scatter.

We stress here that without the color correction described in the previous
section the dispersion of residuals
would be even smaller (by a negligible quantity): 0.052 mag instead of 0.054 mag. 
Therefore, by keeping the data as they are, the data themselves point out that
the intrinsic scatter of the red sequence is very small and not artificially
reduced by our ``corrections" (that instead make it marginally larger!). 

We emphasize that the same result is obtained in two
independent ways: a) by adopting the CWW spectrum and the color
correction (that can be questioned) outlined in the previous section  or, b), 
by adopting a pure empirical
approach (the Neural Network interpolation), that avoid any assumption on color gradients and
on the slope of the color--magnitude relation.

The use of a robust measure reduces the impact of outliers (clusters with wrong
redshift assignment), that are naturally discarded. Even better, we can
estimate the number of clusters with wrongly assigned redshifts, under the
assumption that residuals are Gaussian distributed. In the
distribution shown in Fig 5 there are 220 galaxies, out of 253, within $\pm 2
\sigma$. Therefore, assuming a Gaussian distribution, 230.6 (=220/0.954)
galaxies are expected in absence of wrongly assigned redshifts. We have,
instead, 253 galaxies, i.e. 22 more. This difference ($\sim 10$ \% of the
total number of clusters) is an upper limit to wrongly assigned redshifts.

Part of the observed scatter is of observational nature, not intrinsic to variations
from cluster to cluster of the color of the red sequence. There are a few major
sources of scatter: first of all, Poissonian photometric errors on colors (listed in
the EDR--SDSS database) account for 0.023 mag (Table 1). Second, there are further
photometric errors, due, for example, to the variations of the point spread function
over the EDR--SDSS camera (Gunn et al. 1998). For these errors it seems safe to assume
a 0.02 mag error\footnote{After this paper has been submitted, the SDSS team (Abazajian et al. 
2003) estimates that the zero--point varies by 0.02 mag RMS in $g'-r'$, in perfect
agreement with our guess.}. 
Third, the E track is removed with an accuracy of 0.015 mag (Section
3.1). Finally, the velocity dispersion of galaxies in cluster induces
a $0.01/\sqrt{2}$ mag term (Section 2.2). By subtracting off
quadratically these four terms from the observed scatter, we found 0.041 mag (Table 1). 

%Qui dire che gli errori non sono catastrofici come quelli dei foto-z?

\subsection{Redshift dependence of scatter \& biases}

Figure 6 shows the color residuals from the E track,  after splitting the sample in
near ($0.06<z<0.15$) and far ($0.15<z<0.36$). The centers of the two distributions are
almost identical (Table 1), confirming the accuracy of our color corrections. The two
samples have identical dispersions, 0.040 and 0.041 mag, after correction for photometric errors
(see also Table 1). The two histograms computed instead without aperture and color
corrections would differ by a tiny quantity, some 0.010 mag, but the large number of
data points allows an F test to reject the possibility that the two distributions have
the same mean at an uncomfortably high statistical level. In order not to
incorrectly claim that the color of the red sequence evolves in our
redshift range, color corrections must be applied. 
No evidence for a variation in
the color dispersion is also found when color
variations are removed using our Neural Network interpolation, making the
found dispersion constant irrespective of any assumption on color gradients and
on the slope of the color--magnitude relation.

Highest redshift bins are populated by many ``Red Luminous Galaxies". The
distance between the color cut for the target selection and the E track
is about five times the observed scatter in color. Therefore,
the distribution of color residuals is not biased by cutting its blue tail at
five sigma from the center, especially when one realizes that much less than one
cluster would be removed by such a cut. Hence, the small dispersion in the
higher redshift bin is not due to a selection effect (a correlation between the
galaxy color and the way galaxies are selected for spectroscopic observations),
confirming the conclusions presented in Sect 2.3.

\subsection{Hierarchical scenario}

Most of our sample is composed by low mass clusters and groups, so poor
that they are not listed in the Abell (1957) catalog, not even in the $R=0$
class. In the Abell catalog there are several examples of clusters less
massive than 2 $10^{13} M_\odot$ (see Table 3 in Girardi et al. 1998), and,
therefore, there is no doubt that our sample, that contains many clusters
too poor to be listed in the Abell catalog, includes the groups considered
by Kauffmann (1996). Our sample also includes rich clusters, both because in our
sample there are clusters listed in Abell (1957) with $R>0$, and because
sampling a volume larger than Abell (1957) we have rich clusters not listed
by Abell because too far. Hence, our mass range is appropriate for the
comparison with hierarchical models. 

A solid prediction of the hierarchical scenario is that giant ellipticals in
groups (defined as $10^{13} M_\odot$) are about 4 Gyr younger than giant
ellipticals in clusters (defined as $M \geq 10^{14} M_\odot$). 
 An attentive
inspection of Figure 2 in Kauffmann (1996) clarifies that the claimed 4 Gyr age 
difference holds for the whole magnitude range studied in this paper.
At the midpoint
of our sample, $z=0.2$, a 4 Gyr age difference produces a 0.23 mag difference
in $g'-r'$, adopting the same prescriptions as described in Kauffmann (1996;
i.e. a 11 vs 7 Gyr GISSEL (Bruzual \& Charlot 1993) template). At the highest redshift
end, the difference would be even larger, because of the reduced time between the
start of the star formation and the epoch (redshift) of the observed clusters.

%Therefore, the hierarchical scenario predicts that the red sequence of our clusters spreads
%over a 0.23 mag color range, 

However, the observed dispersion is 0.054 mag (Table 1), and
the intrinsic scatter across clusters $\sim 0.02$ mag (see Section 3.6). No blue tail, 
due to groups and poor
clusters, is seen, contrary to the predictions of the hierarchical model, and in good
agreement with the environmental independence of the Fundamental Plane (Pahre, de Carvalho,
\& Djorgovski  1998; Kochanek et al. 2000). Furthermore, in the next section, we will show
that the color of the red sequence does not correlate with richness, as instead the hierarchical
scenario predicts.

\subsection{Environmental dependence of scatter \& Hierarchical scenario}

Figure 7 shows the color residuals from the E track, cutting the sample 
inside ($R<250$ kpc) and outside ($R>250$ kpc) the cluster core. Red
sequence colors outside the cluster core are bluer, on average, by 0.023
mag, and have a larger (photometrically corrected) dispersion, 0.052 vs
0.030 mag (Table 1).  We interpret the bluer  color and larger dispersion at
larger radius as due to the morphological segregation: as the clustercentric
radius increases, the fraction of spiral galaxies increases. Since spirals
are bluer than early--type galaxies, the color of the red sequence become
bluer and more dispersed at larger clustercentric radii, as observed in 11
clusters by Pimbblet et  al. (2002). It should be mentioned here that many
spiral galaxies in nearby clusters have colors similar to early--type
galaxies: see, for example, the color distribution of the morphological
types in Fig. 3a in Andreon (1996) for a complete sample of galaxies in
Coma, or Table 2 in Oemler (1992). Therefore, many spirals can be within the
color strip centered on the red sequence, but preferentially in the bluer
part, hence making the average color bluer, and the dispersion larger. 

The blueing of the red sequence can be, alternatively, due to an intrinsic
blueing of the early--type galaxies at large clustercentric radii, as
advocated by Abraham et al (1996). The issue cannot be solved with EDR--SDSS
data, because the angular resolution of the imaging data is not adequate to
determine morphological types  with good accuracy. In fact,  a good
morphological  classification needs the following angular resolutions:
$0.06/z$ arcsec (1.9 kpc, $H_0=50$ km s$^{-1}$ Mpc$^{-1}$),  according to
Dressler (1980), 0.8 kpc (Andreon 1996 and Andreon \& Davoust 1997, $H_0=50$
km s$^{-1}$ Mpc$^{-1}$), or even 0.65 kpc, in order to match the high
resolution observations of the {\it Hubble Space Telescope} at $z\sim0.4$
(Andreon, Davoust \& Heim 1997, $H_0=50$ km s$^{-1}$ Mpc$^{-1}$). In angular
terms, this implies sub-arcsec seeing for all clusters in our sample, i.e. a
resolution outside the EDR--SDSS capabilities.
In order to claim that the blueing is really intrinsic, one should also
check if residual interlopers, left over after our spectroscopic selection,
have a role in the observed blueing and increasing scatter at large
clustercentric distances.

The found blueing is related to the {\it proper} distance (i.e. in Mpc)
from the cluster center (or any quantity correlated to it, such as the
local density of galaxies). In fact, if we split the sample in two parts by
considering separately galaxies detected in the inner half detection area
from those detected in the outer half, no differences are found in the
median color or dispersion (Table 1, Figure 8). The latter splitting, while
putting galaxies sitting at larger clustercentric distances preferentially in the
outer sample, mixes together near and far galaxies when proper distances are
adopted, because the detection area is not fixed in Mpc$^2$. In Figure 8 
it could happen that a galaxy with $R<0.25$ Mpc (say) is put in the
outer sample, while another one at larger radii is put in the inner sample. 

Does the red sequence color depends on the cluster richness, as advocated
by the hierarchical scenario? At the present time, we don't dispose of a
good (i.e. redshift invariant)
measure of richness for the detected clusters, but just of the number
of galaxies $N_g$ brighter than $r=21.5$ mag in the detection area.
Depending on redshift, different portions of the luminosity function and
of the cluster area (in Mpc$^2$) are considered. Furthermore, even at a
fixed redshift, a fixed detection area correspond to different cluster
portions. Finally, clusters are best detected on different scales, even at
a fixed redshift, making the situation even more complex. 
Therefore, a clean separation of rich and poor clusters (say richer and
poorer than a given class) is not presently possible. We can, however,
approximately split the sample in two sub--samples whose richness
distributions overlap
somewhat, but whose average richnesses are different. Let us consider 
for simplicity only
clusters best detected at the smaller detection area (that are the
majority), 2.5 arcmin of size. At each fixed redshift, we took a
threshold that divides the sample approximately by two (high and low
richnesses), thus creating two samples whose average richnesses are different
on average. We call them ``dense" and ``loose" clusters, in order to
emphasize that our richness measure is more similar to a central density
than to a global richness. In practice, the observed richness, $N_g$,
depends on redshift approximately as $(1+z)^{-1.5}$, and therefore a
single cut at $11/(1+z)^{-1.5}$ galaxies inside the detection area 
suffices for our aims.

Figure 9 shows the color residuals for the ``loose" (open points) and
``dense" (close points) samples. In rich clusters the red sequence is
redder, on average, than in poor clusters, by 0.015 mag (Table 1). While the
difference is statistically significant (an F test rejects the hypothesis
that the two distributions have the same mean at 99.99 \% confidence
level), the statistical difference should not be, however, overemphasized. First, the
difference is tiny in absolute terms. Second, cluster samples are smaller
than in the previous splitting, because we are obliged to discard clusters
with best detection radii on scales different from 2.5 arcmin. Third, poor
clusters have a larger fraction of spirals, and therefore are more
sensible, than rich clusters, to contamination by bluer galaxies, that
skews toward the blue the color of the red sequence. Fourth, poor clusters are 
(expected to be)
smaller on average than rich ones. Therefore a fixed scale is sampling a portion of
the cluster that is larger in poor clusters than in rich ones, thus again
biasing toward the blue the color of the red sequence of poor clusters.
On the other end, a cleaner division of clusters in rich and poor ones
would probably makes the difference larger, because in this case the
two classes no longer overlap. 

The hierarchical scenario predicts that the brightest elliptical  galaxies are 0.23
mag bluer in $g'-r'$ in groups than in rich clusters, a difference that it
is not seen. A caveat should however be exercised: the computed hierarchical
expectation assumes a metallicity independently on environment. 
If there is a age--metallicity conspiracy  such
that in low--density regions early--type galaxies are younger and also more metal-rich 
than in high--density environments, as suggested by some observations (Kuntschner et al. 
2002), then the two effects may cancel each other  
in the galaxy colors. Therefore, the observed disagreement between theory
and observations may be recovered
if future modeling of the star formation in hierarchical scenarios
will provide the required tuning between age and metallicity.

Finally, it could be argued that some clusters may be contaminated by (possibly infalling)
galaxy groups or composed by filaments/superclusters seen along their major axis. This
hypothetical contamination of other structures in the class of rich clusters does not help
to save the hierarchical scenario, because contaminated clusters should display, in the
hierarchical scenario, a color spread that is not observed.

\subsection{Intercluster/intracluster scatter}

The photometric--corrected observed dispersion is 0.04 mag, or less.
Such a scatter is one of the smallest ever quoted, to the author best knowledge. 
It is composed by two parts: the intercluster scatter, related to
the scatter around the color--magnitude relation, and the scatter of the color
of the red sequence across clusters (intracluster scatter). 

The intracluster scatter has a maximal value of 0.030 mag, as derived
from the photometric--corrected observed dispersion in the cluster core (0.030 mag)
assuming a minimal (zero) intercluster scatter (Table 1). 

The minimal intracluster scatter can be derived
assuming a maximal intercluster scatter (0.038 mag, i.e. the dispersion of an uniform 
deviate between -0.05 and +0.05): 0.018 mag in the whole sample and an imaginary value 
(but compatible with 0.000 mag) in the cluster core. 

Another route is to assume for intercluster scatter
the value observed by Bower, Lucey \& Ellis (1992) for Coma and Virgo, 0.02 mag in $B-V$
(that approximately match $g'-r'$ at the median $z$ of the sample), converted from
$U-V$ using Worthey (1994). Under this assumption, the intracluster scatter is 0.035 mag
and 0.022 mag for the whole sample and for the cluster core, respectively,
still among of the smallest ever quoted.
The latter estimate assumes that poor clusters and groups have red sequences as tight as
rich clusters, which is unlikely given the larger fraction of late--type galaxies in
these environments, part of which fall within the $\pm 0.05$ mag color strip. Therefore,
the latter two figures are upper limits to the scatter across clusters, more than measures. 
This conclusion is reinforced by  Stanford et al. (1998) results: the intercluster
scatter, 0.06 mag (derived for the $red-blue$ color), observed by Stanford et al. (1998) is
systematically a bit larger than, but equal within the (larger) errors of Stanford et al.
(1998), our total one, 0.042 mag (intercluster
plus intracluster). The larger systematic scatter of Stanford et al. suggests
that either these authors underestimate some potentially important sources of
observational scatter (that should be removed from their 0.06 mag scatter), or that it
exists some clusters whose red sequence is more scattered than in Coma and Virgo.

In summary, the direct measure of the maximal and minimal intracluster scatter are 0.030 mag
and zero, respectively. By adopting reasonable, but unmeasured for our sample, assumptions
on the intercluster scatter, the maximal intracluster scatter is $\sim 0.02$ mag.

\subsection{Red sequence color as standard candle}

If the color of the red sequence does not depend on redshift and environment,
it can be used as a distance indicator.

How well can be derived the redshift from the color of the red sequence? The color of
the red sequence cannot be defined for all clusters in the same way as we did for the
subsample of clusters studied in this paper (that have a known redshift). We use
instead the color at which the cluster is best detected. From that color we derive a
$z_{photom}$ assuming the color tracks shown in Figure 4. Both the Coleman, Wu \&
Weedman (1980) E spectrum and a Neural Network interpolation trained on the pairs
($g'-r',z$) give almost identical results. The scatter between $z_{photom}$ and
$z_{spect}$ turns out to be 0.018 in $z$ between $0.06<z_{photom}<0.30$, as shown in
Figure 10. The considered redshift range is a bit reduced than in the remaining of the
paper,  because the flattening of the color--redshift relation (Figure 3) does not
allow to accurately measure $z_{photom}$ at $z>0.3$. The observed scatter, computed on
about 140 clusters with $z_{spect}<0.3$,  is 40 \% smaller than observed in Gladders \& Yee
(2000) at $z<0.5$  for their sample of about 20 clusters, and outperforms by a factor
of 3 all photometric redshift estimates published thus far (except neural redshifts:
Tagliaferri, Longo, Andreon et al. 2003a,b) and by a factor two the
precision of the Fundamental Plane for a single galaxy (0.03, i.e. 15 \% in $z$ at the  
median redshift, e.g. Jorgensen, Franx, \& Kjaergaard 1996). 

The greater efficiency
of the red sequence color as redshift estimator with respect to photometric redshifts 
is likely due to the implicit selection of one single type of galaxies with a
distinctive 4000 \AA \ break (spectrophotometric bright early--type galaxies), more than
the collective use of many galaxies. In fact, the same precision is achieved reducing to
one the number of
galaxies used per cluster (by taking the galaxy defining the color of the red sequence).

We re--emphasize here that our cluster detection algorithm is not biased against
red sequence of any given color, and  the low scatter found is not due to having discarded in the
detection phase clusters that have a red sequence with a color different from the average.
Therefore, the good agreement between $z_{spect}$ and $z_{photom}$ is not the result of
having discarded (undetected) clusters for which such relation does not hold. 
We remind, however, that clusters with a flat color distribution, if they exist,
are under--represented in our sample, because their detection probability is lower than
the one for clusters with a peaked color distribution.

The color--redshift relation used to derive redshift from color is purely empirical,
especially the one adopting the Neural Network
interpolation. Being the relationship empirical, it is independent on the
cosmological model, color corrections, galaxy evolution or k--corrections (and so on).  
Because empirical,
the found relationship cannot be used outside the range in which it is
validated (i.e. extrapolation is not allowed), and a calibration set of pairs of $(color,redshift)$
is need.

\section{Summary of the observational results and discussion of the observational
results}

We studied the color of the red sequence of 158 clusters  detected
by our cluster detection method. They are the clusters with available  $z$ in the
spectroscopic EDR--SDSS database (in the sense explained in section 2). All along the
paper  we show that there is no reason why the subsample of studied clusters
should be different in their color properties from the larger sample of clusters
detected in the EDR--SDSS in the same redshift range. Therefore, our sample of
158 clusters is representative of the whole cluster population detected on the
EDR--SDSS and in the $0.06<z<0.34$ range bare for clusters so rare that their
inclusion in our sample is unlikely. Our results are therefore general because based
on a large sample, and because based on a random subsample of a $\sim 4$ times larger sample 
of clusters. 

All cluster redshifts are based on one single redshift measure. The impact of
wrong redshift assignations (due to interlopers) is reduced by adopting robust
statistics, much less sensible to outliers than the usual mean and standard deviation.

The average color of the red sequence changes, as a function of redshift, as
expected for a red sequence populated by not evolving ellipticals with the
spectrum listed in Coleman, Wu \& Weedman (1980). However, the
leverage given by the studied redshift range is small for discriminating
plausible evolutions, such as a no--evolution from a passive one.

The total measured scatter in color of the red sequences is 0.054 mag. A virtually
identical result is found without any correction for aperture and luminosity effects.
Therefore, the homogeneity across clusters of the color of the red sequence is intrinsic to
the data, not produced by some sly operations performed on data. We emphasize that the same
result is obtained adopting the CWW spectrum plus color corrections or a pure empirical
approach (the Neural Network interpolation), that avoid any assumption on color gradients
and on the slope of the color--magnitude relation.

Clusters detected by our method with a red sequence bluer than the color of an E (in
their rest frame) are not discarded ab initio, but are detected, if they exist. Figure
6 shows the rarity of such a type of clusters.

The overall picture is the one of extreme similarity in the color of red sequence galaxies,
across clusters, and from rich clusters to groups. In itself, it is not a surprise: it is
the level of uniformity and universality that stands out. It holds for an unbiased (in
color) subsample of all detected clusters and groups. A hierarchical scenario would predict
a 4 Gyr age difference between clusters and 10$^{13} M_\odot$ groups, or a 0.23 mag color
difference at the median redshift, still we don't found any group whose red sequence is
more than 0.1 mag bluer than an old elliptical galaxy.  In the hierarchical  scenario, the
galaxy evolution should therefore pushed to high redshift, or, in alternative, high density
regions could exist in a wide range of  environments, and in particular, in the seeds of 
present day groups. One more possibility is that metallicity differences compensate
age differences.

The level of the color uniformity is also outstanding: the observed scatter, inclusive of
observational errors, is about 0.05 mag, the precise value depending on which sample is
considered (see Table 1). Once observational effects are removed the
scatter drops to 0.04 mag. If only galaxies in the cluster core are considered, the
observed scatter of the color of the red sequence is even smaller,  
0.03 mag (observational--corrected), suggesting that the galaxies in the
cluster core are less scattered around the color--magnitude relation than galaxies at
larger clustercentric radii, probably due to the lower fraction of late type galaxies
that reside in the cluster core. 

The color of red sequence measured outside the cluster core is bluer
by 0.023 mag, and have a larger dispersion, by 0.022 mag. We interpret these differences as
due to (red) spirals, whose abundance, relative to early--type galaxies, increases with
clustercentric distance. However, the effect can be alternative interpreted as genuinely
related to early--type galaxies, although indirect observations disfavor such
possibility. Sub--arcsec resolution images are need for definitively
discriminate between the two possibilities.

If clusters are divided in ``dense" and ``loose" with a ``fuzzy" splitting 
that produces two samples of different average richnesses, but with the two classes
overlapping somewhat, then loose clusters have marginally bluer red sequences, and
we ascribe the effect to the increased spiral contamination in these clusters,
with the same caveat as before.

The direct measure of the maximal and minimal intracluster scatter are 0.030 mag and
zero, respectively. By adopting reasonable, but unmeasured for our sample, assumptions on
the intercluster scatter, the maximal intracluster scatter is 0.022 mag.

A few caveats should be emphasized, before to proceed to a quantitative
interpretation of the observational result via some model assumptions.

First: by examining the color distribution, after removal of outliers, we are
describing the {\it average} properties of the red sequence of clusters. The
evidence of a similarity in color for most of the sample does not exclude that
very few clusters had different star formation histories, for example forming
stars later/former than the bulk of the clusters. Figure 5 shows that if these
clusters exists they are very rare. 

Second: we found an homogeneity in color over the whole sampled redshift range.
However, because of the way the EDR--SDSS spectroscopic sample is built, we are
measuring the color dispersion for galaxies in a large (small) magnitude range
at low (high) redshift as explained in section 2.3. Our conclusions cannot be,
therefore, extrapolated to faint galaxies in the highest redshift bins, because
they are not in the sample.

\section{Modeling}

We adopt a simple model to constrain the star formation history of the galaxies on the red
sequence via their color homogeneity across
clusters. The idea is to see what constrains on the
synchronicity or stochasticity of cluster formation are compatible with the small
dispersion (about 0.02 mag intrinsic) of red sequence colors. 
A successful model should produce red sequences with the observed
scatter. We inspired ourselves to Bower, Lucey \& Ellis (1992 and following
works): via a stellar population synthesis model we infer the mean age of the
last episode of star--formation and its scatter by requiring than the expected
color dispersion across clusters matched the observed one. As claimed a few
times in literature, the color evolution is high sensitive to systematic
uncertainties of the stellar population synthesis model, and hence we prefer to
rely only on differential measure only, i.e. on the scatter in color. In our
specific case, systematic errors in the model (Worthey 1992) are larger than the
color differences expected for reasonable choices of the scenarios of star
formation.

%It is worth to mention that the color variation of a GISSEL (Bruzual
%\& Charlot 1993) or Poggianti (1997) 12.75 Gyr old elliptical of solar
%metallicity between $z=0.1$ and $z=0.3$ is short by about 0.1 mag with respect
%to what we observe in $g'-r'$, in the sense that an older galaxy is needed for
%reproducing the observed color variation. However, a 12.75 Gyr old elliptical is
%one that formed their stars at $z=7$ (for $H_0=70$ km s$^{-1}$ km$^{-2}$,
%$\Omega_m=0.3$ and $\Omega_{\Lambda}=0.7$, that we used only for conversions
%from look back time to redshift), and therefore the mismatch in the color
%evolution pushes the formation of E at uncomfortably high redshifts ($z>7$) for
%the presently favored cosmological model, and joins ellipticals to the age
%dilemma associated with globular clusters (Ellis et al. 1997). 

We suppose that the red sequence evolves with a star formation rate (SFR)
exponentially declining (with $\tau=1$ Gyr, a Salpeter Initial Mass Function and a
solar metallicity). Early--type (passive evolving) galaxies have been thus far
modeled in different ways: with a burst of star formation (e.g. Kodama, Arimoto,
Barger \& Aragon--Salamanca 1998), with an exponentially declining SFR (e.g. Bruzual
\& Charlot 1993), with an exponentially declining SFR followed by a truncation (e.g.
Bower, Kodama \& Terlevich 1998), or by a single stellar population (van Dokkum \& Franx
2001). Depending to the reader' preferred way to model passive evolving galaxies, our
choice of SFR can be the one appropriate for a monolithic star formation scenario, or
can account for fresh star formation due to galaxy infall in the cluster or residual
star formation in galaxies on the red sequence. More in general, we are not assuming a
monolithic formation scenario for red sequence galaxies: the exponential decay is
simply intended to characterize the change of the overall rate of star formation,
independently on the fact that galaxies are split between  several subunits or are
single--body objects, and that there is, or not, an infall in the cluster.

If we allow an important secondary infall or a more various star formation
history for the red sequence galaxies, then the constrains we can put on the
last episode of star formation history would be tighter than the one measured.
In fact, any freedom on the evolutionary path increases, not decreases, the
expected dispersion in color, unless some fine tuning is at work. Hence, in
order to keep the color dispersion as small as the observed one, we should force
an higher degree of coordination in the formation of the objects that we now
observe on the red sequence, and/or move the last episode of star formation in
the galaxy far past. It should also be emphasized that we are modeling the {\it
average} evolution of galaxies on the red sequence, not the evolution of any
individual galaxy. Because of the average, the evolution of the ensemble is
surely smoother than the evolution of one single object, hence supporting our
choice of a simple evolutionary model. 

In a given cluster, all galaxies that are {\it now} on the red sequence begin
their star formation all at the same time. The rationale behind this choice is
that we have observationally separated the color spread inside each cluster (due
to an age/metallicity spread) from the color scatter across clusters. The latter is
interpreted as an age spread. The alternative, that the color scatter across
clusters is given by a variation in metallicity from cluster to cluster, is
interesting, but outside the aim of this paper.

While all galaxies (or subunits) inside each cluster begin to form stars at the same
time, $t_{start}$, such a time differs from cluster to cluster. Therefore, each
cluster has its own $t_{start}$. The $z$ distribution of $t_{start}$ is assumed to be
uniform between $z_{start}$ and $z_{end}$.  $z_{start}$ and $z_{end}$ are the
redshift (ages) of the oldest and youngest clusters respectively. More complicate
distributions can be chosen, but we anticipate that $z_{end}$ is the most important
parameter, and that what happens at earlier times is largely unconstraint. Therefore,
there is no need for any supplementary unconstraint parameter.

The distribution of formation redshifts, in spite of his simplicity,
can capture many different {\it global} star formation 
rates: the only common behavior shared by the different global SFR is
that at $z<z_{end}$ the SFR decrease almost linearly (in logarithmic
units) with $z$. 

In order to understand our sensitivity to the library used for the stellar
population templates, we adopted both the GISSEL98 (Bruzual \& Charlot 1993) and
Poggianti (1997) libraries. Since almost identical results are found, we present
results for GISSEL library only, that is sampled with a finer time resolution,
and hence need only smaller interpolations between ages. In order to check our
code, we verify that K and E corrections (whose moments are at the base of our
computations) computed by us are identical to the values listed by Poggianti
(1997) for the same library and filters (we use Johnson filters in the
comparison, for lack of SDSS filters in the Poggianti' set).

In practice, we generate 1000 objects, each one representing the average of all
galaxies on the red sequence of an individual cluster, with the appropriate
$t_{start}$ distribution, and we left the objects to evolve toward the observed redshift,
$z_{obs}$, assuming an exponentially decreasing star formation history. Then, we
(synthetically) observed them through the EDR--SDSS $g'$ and $r'$ filters including the
effect of CCD quantum efficiency, telescope and atmosphere transmission and we compute
their scatter in color at $z_{obs}=0.1$ and $z_{obs}=0.25$. These two redshifts are the
midpoints of the two redshift ranges studied in the previous sections. Results at
$z=0.1$ hold for the color of red sequence, as measured over a 3-4 mag magnitude
range, while results at $z=0.25$ hold only for the brightest galaxies. Such a choice
is dictated by the limitations of the used observational data set: we never measured
the color of distant and intrinsically faint galaxies simply because they are not
spectroscopic targets of the EDR--SDSS.

Figure 11 shows the constrains we can put on the star formation histories from
the scatter of the color of red sequences at $z=0.1$. In this figure, as well as
in the next one, only half of the region is permitted, because the star
formation cannot stop before to start. The dotted line marks the
redshift at which the universe is 1 Gyr older that at $z_{start}$. The elapsed time
between $z_{start}$ and $z_{end}$ is less than 1 Gyr for points above
the curve. An observed intrinsic dispersion of 0.02 mag (middle curve) can be
produced by several pairs of $z_{start}$ and $z_{end}$. If stars in the oldest clusters
formed at $z>2$, i.e. $z_{start}>2$, then the stars in the youngest clusters
formed at $1.1<z_{end}<1.4$, a very precisely determined time for the age of stars
in the youngest clusters (there is about 1 Gyr between $z=1.1$ and $z=1.4$).
If the star formation epoch of the oldest clusters starts at smaller redshift,
say $z_{start}=1$, then the start of the star formation in all the other
clusters should immediately follow, implying an extreme coordination in the star
formation histories of galaxies far apart in the universe and sitting in
different environments (from rich clusters to poor groups), a coordination that
the hierarchical scenario seems not to allow.  If the maximal delay between the
begin of the star formation in massive halos (clusters) and smaller ones
(groups) is  1 Gyr, i.e. 1/4 of the time delay claimed to be by hierarchical
scenarios, then the youngest clusters should form their stars at $z_{end}>1$,
and this holds both for clusters and groups.

While the observed small scatter in color does not exclude by itself that stars in red
sequences galaxies formed at $z<1$ with extreme coordination,  this model is at strong
discrepancy of the mere existence of galaxies at $z=3$ (e.g. Shapley et al. 2001) that
are not allowed if $z_{start}<3$, unless star formation is delayed in clusters,
contrary to what hierarchical scenarios predicts. Furthermore, the model colors would
be too blue to accommodate red sequence galaxies in the redshift range $0.3<z<0.9$
observed by Stanford, Eisenhardt \& Dickinson (1998). Therefore, in order to keep the
dispersion at the observed value and, at the same time, not to contradict the mere
existence of high redshift galaxies, we should discard the possibility of a late and
extremely well synchronized star formation, and keep only the remaining possibility: 
the latest star bust episode in red sequence galaxies happens at
$z\sim1.1-1.4$. Given the fact that the 0.02 mag dispersion in color is probably an
upper limits, then the $z\sim1.1-1.4$ constrain is an lower limit. 

Figure 12 shows the constrain allowed by the color scatter of the red sequence at
$z=0.25$. With respect to the previous Figure, curves move up, showing that the
coordination is tighter. We are able to put stronger constrains for two reasons:
because we observed earlier in time, where galaxies were younger and color differences
had less time to dump out and because we are observing a bluer part of the galaxy 
spectrum, part that is more sensitive to star formation episodes. Star formation stops
at $z\sim2-2.5$, almost independently on $z_{start}$, unless the star formation across
all structures (clusters and groups) is synchronized at better than 1 Gyr. Again, red
sequence colors at high redshift and the existence of galaxies (i.e. star) at high
redshift ($z=3$ at least), both reject the low redshift part of the curve, as for the
previous figure.

Both Figure 11 and 12 shows other two curves: for an intrinsic scatter of 0.04
mag, ruled out by observations, and a 0.01 mag scatter.  The
conclusion are qualitatively the same, but differ quantitatively.
It is worth mentioning that in Figure 12 the track corresponding to an observed
scatter of 0.01 mag runs almost parallel to the $z_{start}+1$ Gyr track. This means
that the elapsed time between $z_{start}$ and $z_{end}$ is almost constant,
independently on $z_{start}$. Therefore, if the observed scatter 
of bright galaxies on the red sequences is 0.01 mag, then
the start of the star formation is extended and coordinated over a very short 
period, 0.5-0.8 Gyr only, shorter than the typical time (1 Gyr) adopted for
describing the star formation in each clusters. I.e. coordination across clusters
would be more precise than inside clusters.

\section{Conclusions}

We studied the homogeneity, across clusters, of the color of the red sequence (the
intercept of the color--magnitude relation) of 158 clusters and groups detected in the
Early Data Release of the Sloan Digital Sky Survey (EDR--SDSS) in the redshift range
$0.06<z<0.34$. Because of the method used to detect cluster, 
clusters with a flat color distribution, if they exist,
are under--represented in our sample.

We found a high degree of homogeneity: the color of the red sequence
shows an intrinsic scatter of 0.02 mag across clusters, suggesting that either
galaxies on the red sequence formed a long time ago ($z>2$) or else their star
formation is universally delayed with preservation of a small spread in formation epoch.
The latter possibility is ruled out by the mere existence of galaxies at high
redshift. While the old age of early--type galaxies was already been claimed for a small
heterogeneous collection of clusters, most of which are rich ones, we found that it
holds for ten to one hundred large sample, representative of all clusters and groups
detected on the EDR--SDSS. Hence we claim the possible universality of the color of
the galaxies on the red sequence at $0.06<z<0.34$. Furthermore, the sample includes a large number of
very poor clusters (also called groups), not studied in previous works, for which the
hierarchical and monolithic scenarios of elliptical formation predict different colors
for the brightest ellipticals. The observed red sequence color does not depends on
cluster/group richness at a level of 0.02 mag, while a 0.23 mag effect is expected
according to the hierarchical prediction. Therefore, the stellar population of red
sequence galaxies is similar in clusters and groups, in spite of different halo
histories. Finally, since the observed rest--frame color of the red sequence does not
depend on environment and redshift, it can be used as a distance indicator, with an
error $\sigma_z=0.018$, a few time better than the precision achieved by other
photometric redshift estimates and twice better than the precision of the Fundamental
Plane for a single galaxy.

The old age of stars in galaxies on the red sequence is derived by using only the
color scatter, and not the color variation, the former being more robust than the
latter to systematic uncertainties in the models. Furthermore, the found old age is a
lower limit if there is a source of photometric errors in the EDR--SDSS not accounted
for (a concern that in fact is general to all previous similar studies too).
The old
age of the stars in galaxies on the red sequence is in comfortable agreement with
previous similar studies, such as the tightness of the fundamental plane relation for
elliptical in local clusters (Renzini \& Ciotti 1993), the tightness of the
color--magnitude relation up to $z\sim 1$ (references listed in the Introduction), 
and the modest shift with redshift of the fundamental plane and color--magnitude
relations (e.g. van Dokkum, Franx, Kelson, \&  Illingworth 1998; van Dokkum et al.
2000, Stanford et al. 1998, Treu et al. 2001). With respect to most of these studies, 
we used quantities
less susceptible of systematic errors in the models (such as the color variation),
and we consider large and controlled samples, inclusive of groups.

%Finally, under the hypothesis (shared by previous works) that our understanding
%of the photometric errors is good, and that star formation starts early in
%clusters, we tentatively suggest that the star 
%formation of red galaxies in clusters and groups 
%extended over a significant redshift range, in order to
%make the scatter in color as large as we measure.

%It has been argued by van Dokkum \& Franx (2001) that the low scatter in the
%color--magnitude relation, indicative of a high formation redshift of their stars, 
%is instead a consequences of the ``progenitor bias'': the progenitors of 
%the youngest low-redshift, early-type galaxies drop out of the sample at
%high redshift. This concern does not apply to our analysis: the galaxies
%are not morphologically selected

%Second, by tracking the red sequence at different
%redshifts, on does not trace the continuous evolution of a single collection of
%galaxies, but instead selects only the oldest galaxies at any given redshift,
%which thus mimic a passive evolutionary behavior or even a not evolving sample 
%(Stanford, Eisenhardt \& Dickinson 1998; van Dokkum \& Franx 2001). Non so

\begin{acknowledgements} 

%During this work, I discovered that galaxies are not the uniq entities
%in the universe that need cosmological times to evolve. Many thing,
%much more near, need similar times.

We acknowledge the referee, T. Kodama, for his comments that
significatively help in clarifying the cluster detectability.
It is a pleasure to acknowledge discussion with D. Burstein, R. de Carvalho, G. Carraro,
R. Ellis, J.-M. Miralles and T. Treu.  Partial founding from ESO, ASI and from the 
conveners of the
WS-GE (through project PESO/PRO/15130/1999 from FCT/Portugal) is acknowledged. 
M. Massarotti kindly provide us
the GISSEL elliptical spectra. A special thank goes to the EDR--SDSS collaboration for
their excellent work, and to Alfred P. Sloan for his gift to the astronomy. The standard
acknowledgement for researches based on EDR--SDSS data can be found in  the EDR--SDSS Web
site: http://www.sdss.org/.

\end{acknowledgements}

\vfill\eject

%\newpage

\newpage

\begin{table*}
\caption{The data}
\begin{tabular}{lrrrrrr}
\hline
{Sample} & {$\Delta$} & {$\sigma$} & {$N_{gal}$} & {$N_{clus}$} & {$\sigma_{photom}$} & {$\sigma_{photocorr}$}\\
\hline 
$0.06<z<0.36$ 	& -0.001 & 0.054 & 253 & 158 & 0.023 & 0.041 \\
$0.06<z<0.15$ 	& -0.002 & 0.051 & 128 & 55  & 0.018 & 0.040 \\
$0.15<z<0.36$ 	& +0.001 & 0.060 & 125 & 103 & 0.035 & 0.041 \\
$R<250$ kpc 	& +0.005 & 0.045 & 126 & 89 & 0.021 & 0.030 \\
$R>250$ kpc 	& -0.018 & 0.063 & 127 & 69 & 0.024 & 0.052 \\
inner half	& +0.000 & 0.057 & 111 & 95 & 0.025 & 0.044 \\
outer half	& -0.002 & 0.055 & 142 & 63 & 0.022 & 0.043 \\ 
half poorer, $l/2=1.25$ arcmin & +0.001 & 0.042 & 62 & 41 & 0.032 & 0.008 \\
half richer, $l/2=1.25$ arcmin & +0.016 & 0.050 & 84 & 58 & 0.025 & 0.035 \\
\hline 
\end{tabular}
\hfill \break
$\Delta$, $\sigma$ are the offset, relative to the E track, and the dispersion,
respectively.
$N_{gal}$ snd $\sigma_{photom}$ are the number of galaxies in the sample
and the dispersion due to Poissonian photometric errors.
$\sigma_{photocorr}$ is the scatter corrected by all photometric errors: 
Poissonian photometric errors, differences of the PSF in the camera, accuracy of the
color trend removal, velocity dispersion of clusters.
\end{table*} 
\newpage 
\null 
\newpage 

\newpage

\begin{figure*}
\psfig{figure=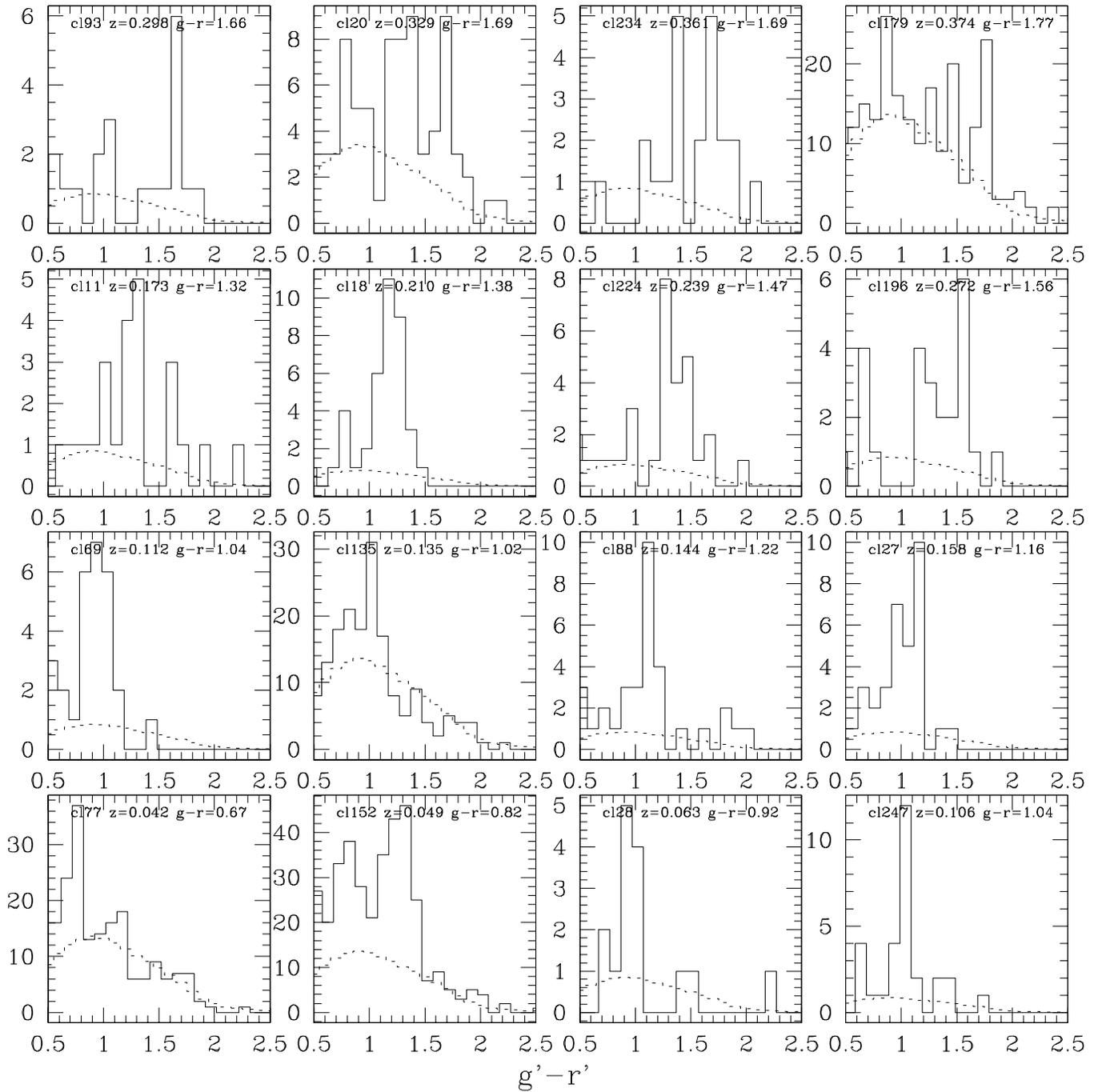,width=18truecm}
\caption[h]{Color distribution in the cluster line of sight
(solid histogram) and on 100 deg$^2$ (dashed histogram), normalized
to the cluster detection area.
}
\end{figure*}

\begin{figure*}
\psfig{figure=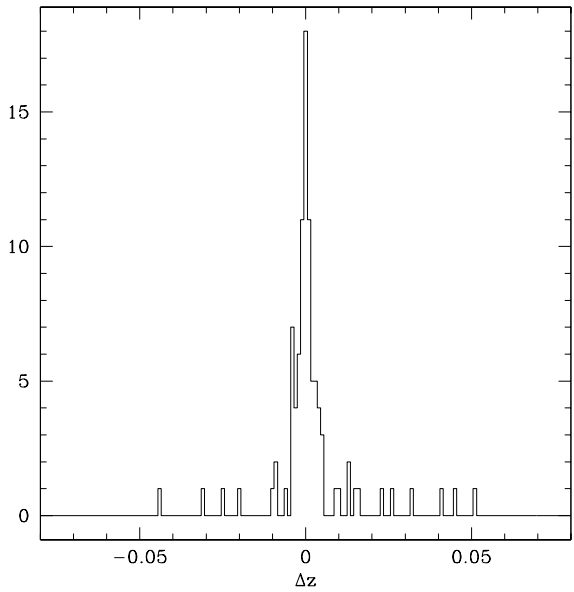,height=8truecm}
\caption[h]{Histogram of the $z$ differences for clusters having two or
more galaxies with redshift.}
\end{figure*}

\begin{figure*}
\psfig{figure=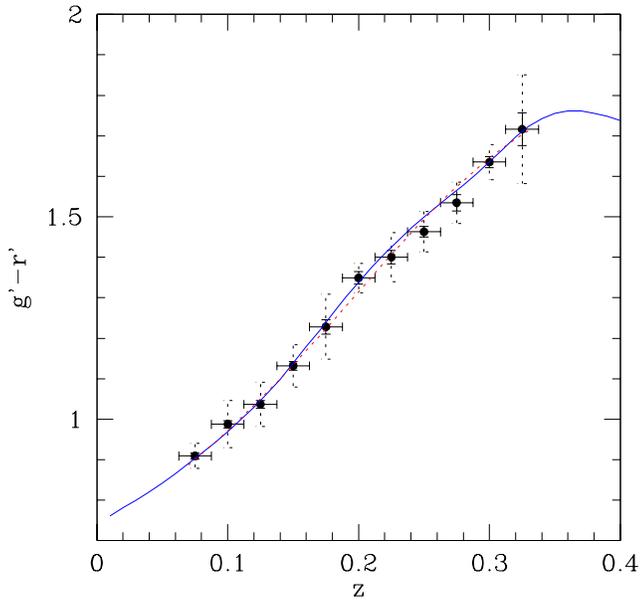,height=8truecm}
\caption[h]{Observed $g'-r'$ color of the galaxies on the red sequence
as a function of z. Dotted error bars in the ordinate mark the observed scatter
around the median, while solid error bars mark error on the average.
Error bars in the abscissa mark half the bin width.
The solid curve is the expected location of an unevolving E, while the
dotted line (barely different from the solid line) is an empirical
interpolation. There are 15 clusters per point, on average.}
\end{figure*}

\begin{figure*}
\psfig{figure=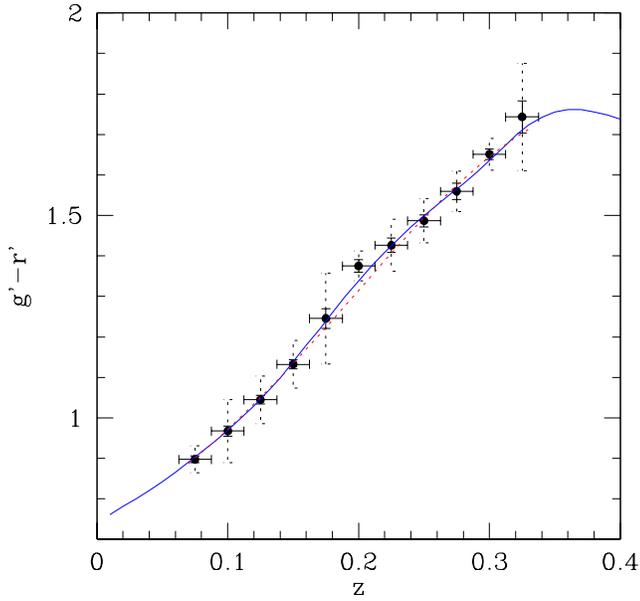,height=8truecm}
\caption[h]{Observed $g'-r'$ color, corrected for the color--magnitude
relation and the color gradients inside galaxies,
of the galaxies on the red sequence
as a function of $z$. Error bars and curves are as in Figure 2.
}
\end{figure*}

\begin{figure*}
\psfig{figure=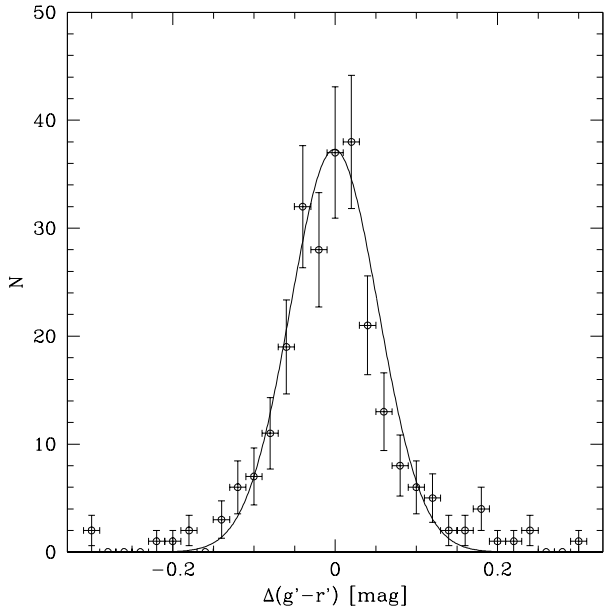,height=8truecm}
\caption[h]{Distribution of color residuals from the expected locus of an unevolving E. The
curve is a Gaussian of the same median and (robust) dispersion of the data,
traced to guide the eye, but never used in the paper.
Error bars on data points are $\sqrt{N}$ (in the ordinate direction) and
half the bin width (in the abscissa direction)
}
\end{figure*}

\begin{figure*}
\psfig{figure=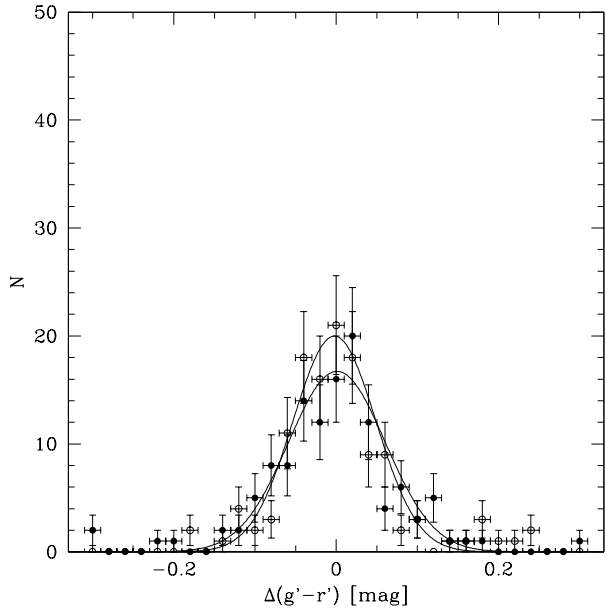,height=8truecm}
\caption[h]{Distribution of color residuals from the expected locus of an unevolving E,
for nearby clusters ($0.06<z<0.15$, open dots) and 
distant clusters ($0.15<z<0.34$, close dots). 
Curves and error bars as in the previous Figure.
}
\end{figure*}

\begin{figure*}
\psfig{figure=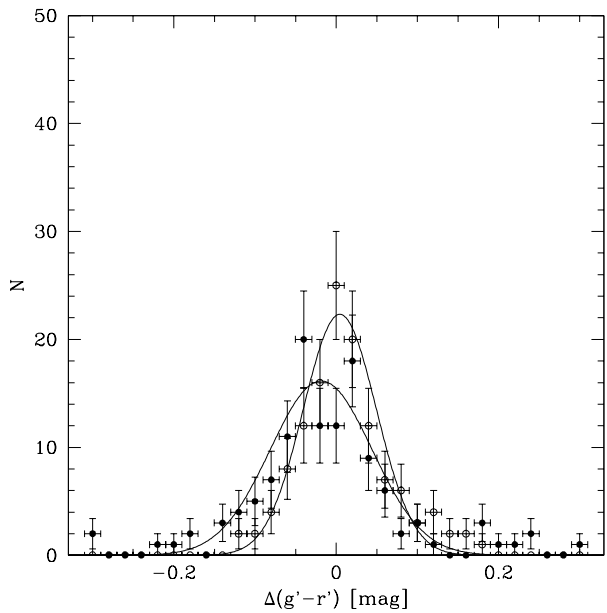,height=8truecm}
\caption[h]{Distribution of color residuals from the expected locus of an unevolving E,
for galaxies inside ($R<250$ kpc, open dots) and 
outside ($R>250$ kpc, close dots) the cluster core. 
Curves and error bars as in the previous Figure.
}
\end{figure*}

\begin{figure*}
\psfig{figure=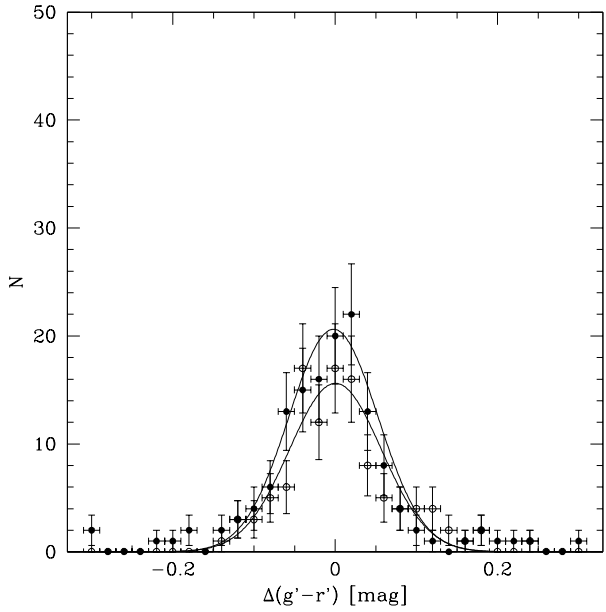,height=8truecm}
\caption[h]{Distribution of color residuals from the expected locus of an unevolving E,
for galaxies in the inner half of the cluster detection area
(open dots) and in the outer half detection area (close dots). 
Curves and error bars as in the previous Figure.
}
\end{figure*}

\begin{figure*}
\psfig{figure=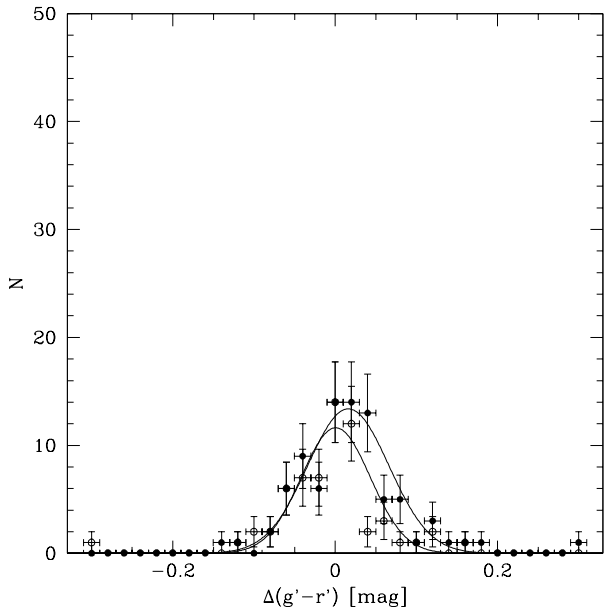,height=8truecm}
\caption[h]{Distribution of color residuals from the expected locus of an unevolving E,
for ``dense" (solid dots) and ``loose" (open dots). 
Curves and error bars as in the previous Figure. See text for the definition
of dense and loose clusters.
}
\end{figure*}

\begin{figure*}
\psfig{figure=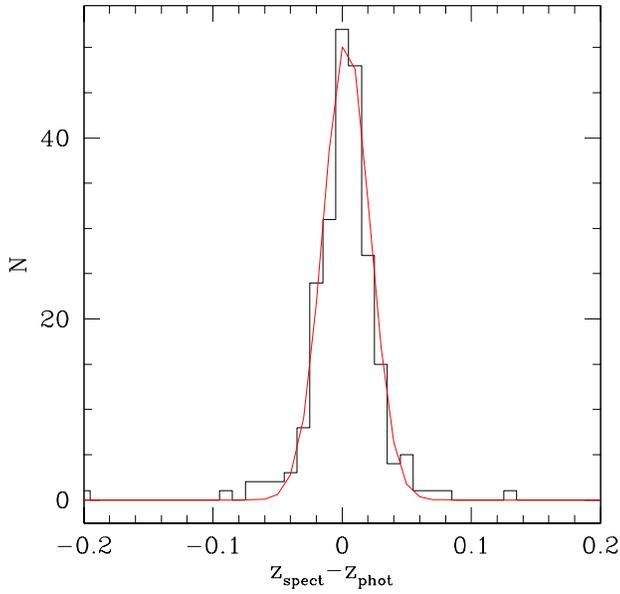,height=8truecm}
\caption[h]{Residual between photometric and spectroscopic
redshifts. The curve shows a Gaussian of the with of the observed distribution
and of the same area. }
\end{figure*}

\begin{figure*}
\psfig{figure=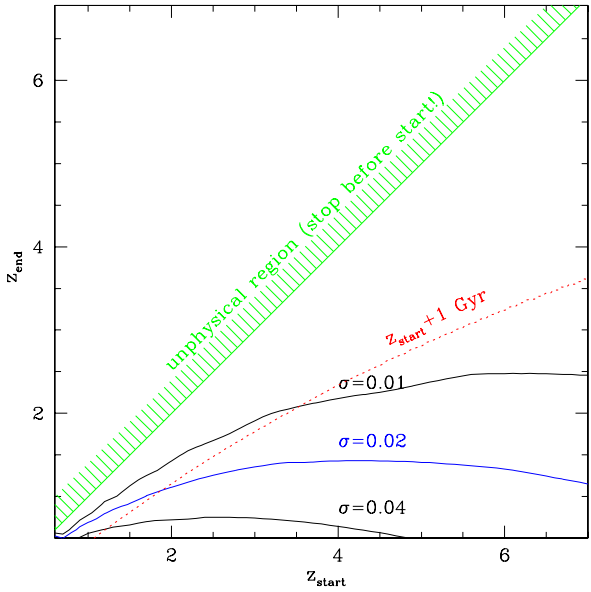,height=8truecm}
\caption[h]{Constrains on the star formation history of galaxies on the
red sequence at $z=0.1$, for three values of the intracluster
color dispersion: 0.04, 0.02, and 0.01 mag.  The dotted line marks the
redshift at which the universe is 1 Gyr older that at $z_{start}$.}
\end{figure*}

\begin{figure*}
\psfig{figure=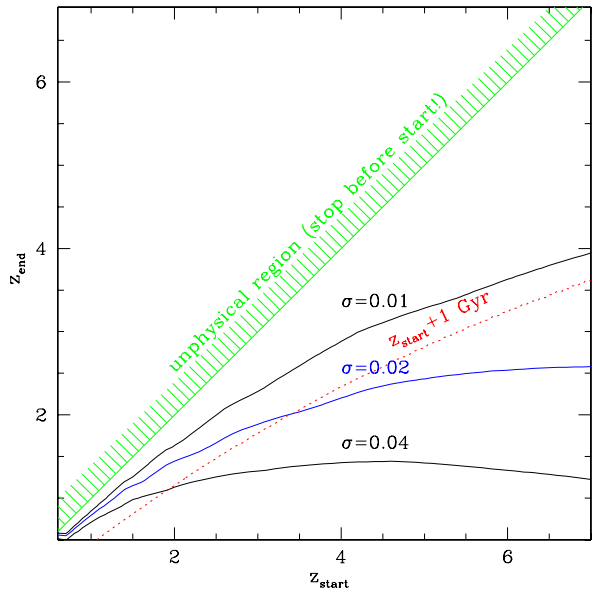,height=8truecm}
\caption[h]{As previous figure, but for $z\sim0.25$ galaxies.}
\end{figure*}

\end{document}